# The effect of aperture and seeing on the visibility of sunspots when using early modern and modern telescopes of modest size

Nicolàs de Hilster[1] and Siebren van der Werf[2]

**Abstract:** Using the convolution of seeing and diffraction, the relation between seeing and aperture in the visibility of sunspots is explored. It is shown that even telescopes with apertures smaller than 5 centimetres are significantly affected by seeing. Although larger aperture instruments suffer more from seeing than smaller ones, their level of detail always remains better under the same atmospheric conditions.

**Keywords:** Sunspots – Aperture – Seeing – Galilean Type Telescope

## Introduction

In his 1993 "Visibility of Sunspots" article Bradley E. Schaefer provided us with the maths to calculate the visibility of sunspots with the unaided eye and with telescopes.[3] Schaefer showed that the visibility of sunspots with telescopes depends on the properties of the sunspot themselves, the smearing functions for the stability of the atmosphere (also known as seeing) and the aperture (due to Fraunhofer diffraction). Their combined effect was given as the addition of their second moments (respectively $\mu_{spot}$, $\mu_{seeing}$ and $\mu_{diff}$), but as $\mu_{diff}$ cannot be directly evaluated by an analytical integration, Schaefer had to use an approximation for that, while he stated that "...in practice, it is difficult to calculate the convolution [of the smearing functions] and impossible to present the result in a simple and general manner."[4] One of conclusions of Schaefer's approach was that telescopes with apertures smaller than 5 centimetres will remove the effects of variable seeing.[5]

During his daily sunspot observations for AAVSO's project *Solar Observing Project: Re-calibrating Historical Sunspot Observations* under guidance of Leif Svalgaard,[6] one of the authors (NdH) noticed, however, that even when using a 1490mm focal length Galilean type telescope, which is stopped down to 3 centimetres (f/50), variable seeing causes sunspots to temporarily disappear from the projected view. When imaging the Sun with a modern 150mm f/7 aperture apochromatic refractor, the images appeared to be less affected by seeing than when using a modern 279mm aperture f/10 Schmidt-Cassegrain reflector. The maths provided by Schaefer therefore seemed to be incomplete as seeing and aperture are not correlated in it. A discussion on this topic between the present authors resulted in the question whether the relationship between aperture and seeing could be modelled. Similar work has been done in by Nina V. Karachik, Alexei A. Pevtsov and Yury A. Nagovitsyn in 2019, but related to 3-5% scattered light for various apertures.[7] The present article discusses the effect of seeing in combination with various apertures.

**Convolution of diffraction and seeing**

Light that passes through a round aperture can be modelled using a Fraunhofer diffraction distribution, the normalised intensity of which is given by:

$$I_{Fraunhofer}(x) = \frac{1}{\pi}\left[\frac{J_1(x)}{x}\right]^2 \qquad [1]$$

Where $J_1$ is the Bessel function of the first kind and argument $x = (\pi D/\lambda)\sin(\theta)$. In the latter $D$ is the diameter of the objective, $\lambda$ the wavelength of light. $\theta$ is the angle as measured from the normal to the aperture. In the region of interest $\theta$ will always be small enough to justify the replacement of $\sin(\theta)$ by $\theta$, which then stands in radians. The Fraunhofer diffraction distribution has a central illuminated area, which is known as the Airy-disk, bounded by a first dark ring at $x = 3.83170597$. Secondary maxima beyond that are much fainter.

The angular radius of the Airy-disk $r_{Airy\text{-}disk}$ in radians follows from:

$$r_{Airy\,disc} = \left(\frac{3.83170597}{\pi}\right)\frac{\lambda}{D} = 1.2197\left(\frac{\lambda}{D}\right) \qquad [2]$$

Using $D = 30$mm and $\lambda = 574$nm results in an Airy-disk radius of 2.334 $10^{-5}$ rad = 4.81".
At a distance of $f = 1.490$ m this results in a round spot with a diameter of $2 \times 1.490 \times 2.334\ 10^{-5}$ m $\approx 0.070$ mm (we disregard here the further magnification by an underlying lens and projection system).

Before entering the objective side of the telescope the light from the star is smeared while it passes through the atmosphere, with a typical size of a few arc-seconds. This smearing follows a normal distribution and it is generally referred to as *seeing*. Quantitatively, its value is usually taken as twice the rms-radius of the smeared area.[8] At the observatory of one of the authors in Castricum, the Netherlands, the average seeing typically varies between 1.5 and 7 arc-seconds.[9]

We assume that the smearing function is a radially symmetrical Gaussian distribution with standard deviation $\sigma$. Taking $\theta$ in radians implies adopting a half-sphere of unit radius behind the aperture, and expressing the polar angle $\theta$ and the azimuthal angle $\varphi$ in this frame. Again, since $\theta$ will always be sufficiently small to neglect the difference between $\sin(\theta)$ and $\theta$, we may also adopt a Cartesian frame in a plane tangent to this unit sphere and parallel to the aperture plane. The convolution will be rotationally symmetric and evaluating it in this plane for a chosen distance $r$ from the origin, may be achieved by adopting the $x$-axis parallel to $r$ and the $y$-axis perpendicular to it. The convolution then reads:

$$I_{convolution}(r) = \iint dx\,dy\,\frac{1}{\pi}\frac{\left[J_1(\sqrt{x^2+y^2})\right]^2}{x^2+y^2}\frac{1}{2\pi\sigma^2}\exp\left(\frac{-(r^2-x)^2-y^2}{2\sigma^2}\right) \qquad [3]$$

---

8   Seykora, E.J., "Solar Scintillation and the Monitoring of Solar Seeing", in: *Solar Physics, vol. 145*, (1993), pp.390-91, Also see footnote 19. One may also find seeing defined as its Full Width at Half Maximum (FWHM): FWHM=$\sqrt{(2\ln(2))}\times(2\times$rms-radius$) = 2.355\sigma$, see: ESO, "Analysis of telescope image quality data", url: http://www.eso.org/gen-fac/pubs/astclim/papers/lz-thesis/node57.html (last accessed 12 August 2022), Schaefer op. cit., p.411, Karachik, e.a., op. cit., p.3.

9   As measured by NdH using a *Solar Scintillation Seeing Monitor* in the period September 2021 - August 2022.

In this convolution both the Bessel and the Gauss functions are normalised to 1 in order to have the resulting convolution normalised as well.

What the convolution basically does, is that for each point on the 2D Fraunhofer distribution shape the surrounding area is sampled with a radius given by the Gaussian distribution. The sum of this is then taken as the intensity value for that point.[10]

Figure 1 shows the effect of convoluting a Fraunhofer distribution with a Gaussian of standard deviation $\sigma$. A low value of $\sigma$ affects the Fraunhofer diffraction distribution only marginally, but from about $\sigma = 1.5$ the convolution already starts to become bell-shaped like a Gaussian distribution, and the convolution will no longer be surrounded by brighter rings as in the original Fraunhofer diffraction distribution.[11]

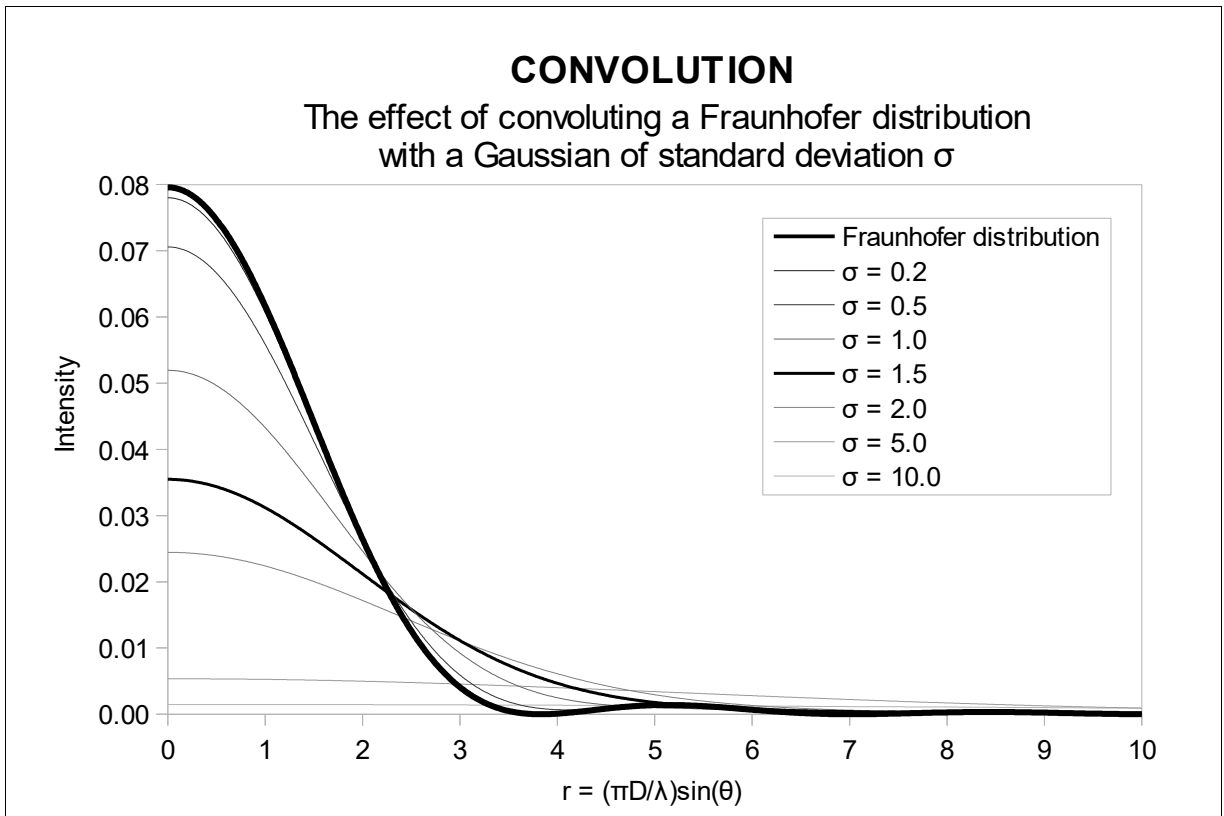

*Figure 1: The effect of convoluting a Fraunhofer distribution with a Gaussian of standard deviation $\sigma$.*

Convoluting the Fraunhofer distribution with the Gaussian seeing distribution requires that their arguments must have a common scaling. As mentioned above, we take seeing as twice the rms-value of the latter function. The conversion from arc-seconds to the scale of the Fraunhofer function, on which the central maximum extends to 3.83170597, reads:

$$\sigma = \frac{seeing}{2} \times \frac{\pi \times D}{\lambda} \times \frac{\pi}{180 \times 3600} = \frac{seeing}{2} \times \frac{3.83170597}{r_{Airy\ disc}} \qquad [4]$$

---

10 The convolution assumes a round aperture without central obstruction. All simulations shown in this article are based on this assumption, while it is also assumed that the optics are perfectly figured and polished.

11 This transition takes place somewhere between $\sigma = 1.0$ and $\sigma = 1.5$.

Here seeing and Airy-disk radius are in arc-seconds and as the value for seeing represents its diameter, it is divided by 2 to get its radius.

For the above mentioned 30 mm aperture telescope $r_{Airy\text{-}disk}$ equals 2.334 $10^{-5}$ rad = 4.81". Suppose we have a seeing of 3", which is the twice the rms-radius of the smeared area, then the standard deviation to be used for the convolution will be $\sigma$ = (3 / 2) × (3.83170597 / 4.81) = 1.19, significantly smaller than the first minimum of the Fraunhofer distribution, making this telescope diffraction-limited.

Suppose we use a telescope with an aperture of 280 mm at a wavelength of 574nm, the Airy-disk will have a $r_{Airy\text{-}disk}$ of 2.50 $10^{-6}$ rad = 0.52" from the optical axis. Suppose we still have a seeing of 3", the standard deviation $\sigma$ to be used for the convolution is (3 / 2) × (3.83170597 / 0.52) = 11.14, significantly larger than the first minimum of the Fraunhofer distribution. In this case the telescope is seeing-limited.

Figure 2 shows the relationship between seeing, aperture and the standard deviation $\sigma$ to be used in the convolution. As the Airy-disk radius is inversely proportional with aperture the corresponding $\sigma$ grows quicker with seeing for larger apertures than for smaller ones, which implies that the image deteriorates quicker in larger aperture telescopes.

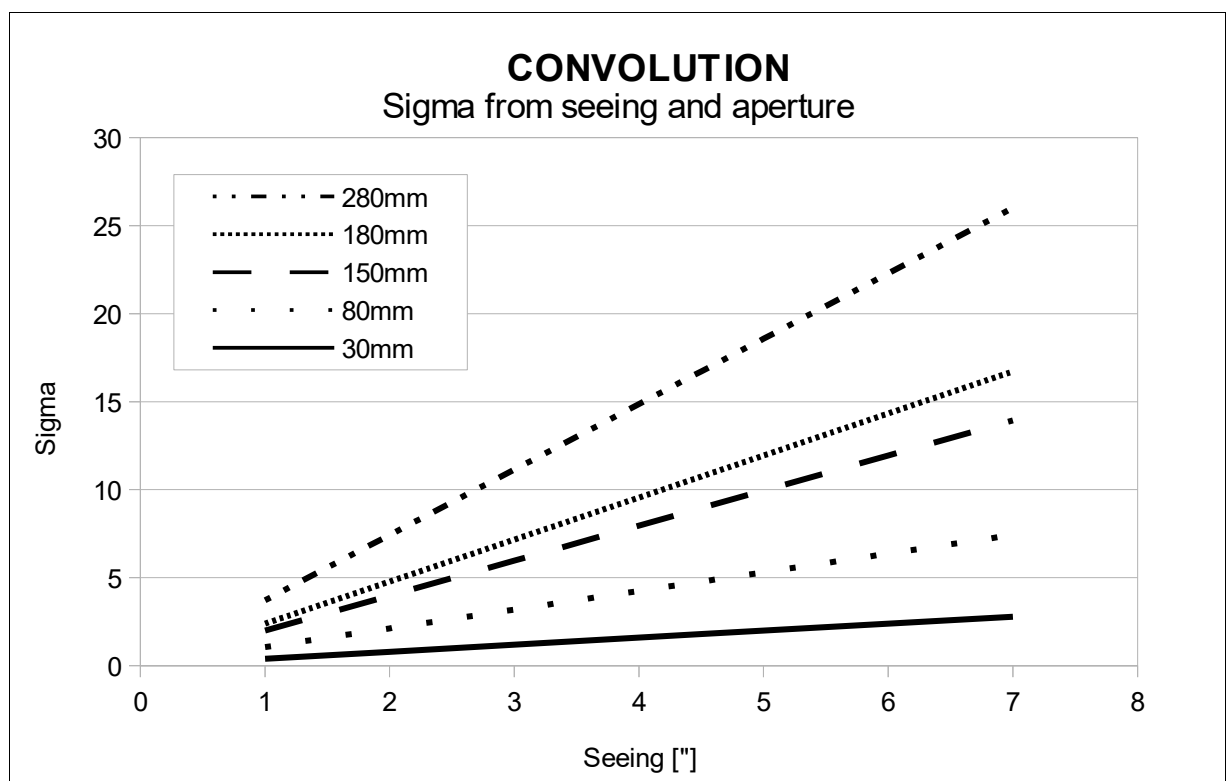

*Figure 2: The relation between aperture and the standard deviation of seeing, σ.*

Further below, graphs show the effect seeing has on angular and metric smearing. In order to produce those graphs a probability level had to be chosen that limits the calculations as both the Fraunhofer and Gaussian distributions, and thus their convolution, have no angular boundaries. Figure 3 shows the effect of $\sigma$ and probability level on the diameter of the illuminated zone in Airy-disk radii.

Five levels have been tested: 39.4% (Gaussian 2D 1σ), 50.0% (Half Flux Radius), 83.8% (Airy-disk radius), 86.5% (Gaussian 2D 2σ), and 98.9% (Gaussian 2D 3σ). That last level clearly causes the illuminated area to grow out of proportion at low σ. As the Half Flux Radius (HFR, 50%) is commonly used to measure star diameters (e.g. in focusing routines) we decided to continue with this value for the following graphs.[12]

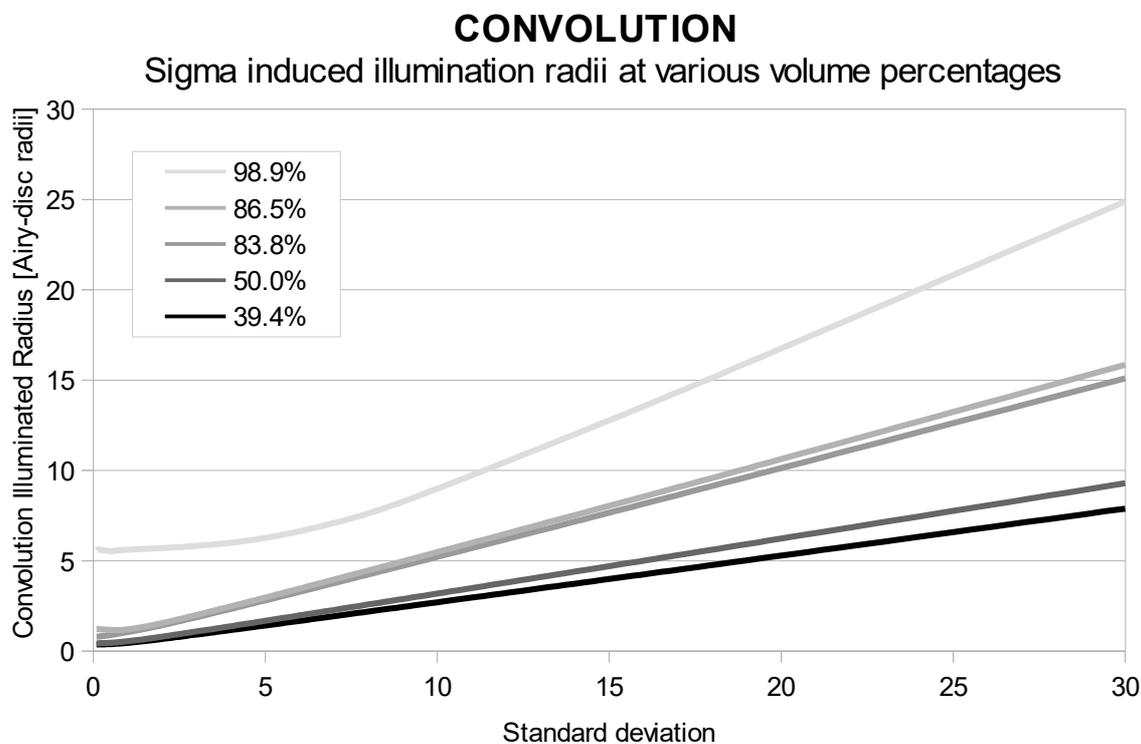

*Figure 3: The relationship between standard deviation σ, the volumetric probability level and diameter of the illuminated circle.*

Larger aperture means a smaller Airy-disk, so figure 4 shows the relationship between actual seeing and the smearing effect due to diffraction and seeing after convolution. The calculations are independent of focal length. From that graph it becomes apparent that, when looking at angular smearing, all scopes with larger aperture than 30mm are more or less equally affected by seeing. Only the 30mm aperture telescope shows a relatively larger convolution radius at lower seeing values due to the fact that this telescope is diffraction limited at those values.

As in practice telescopes do not have uniform focal lengths, the differences will cause images to be produced at different scales. So although the angular smearing remains the same in figure 4, the metric smearing (i.e. the smearing we experience when projecting or imaging a sunspot) will vary as a result of the focal length. Figure 5 shows the metric smearing of the same five telescopes, using their non-uniformal focal lengths.

---

12 The convoluted sunspot images shown further below are produced using a Gaussian 2D 3σ level (98.89%).

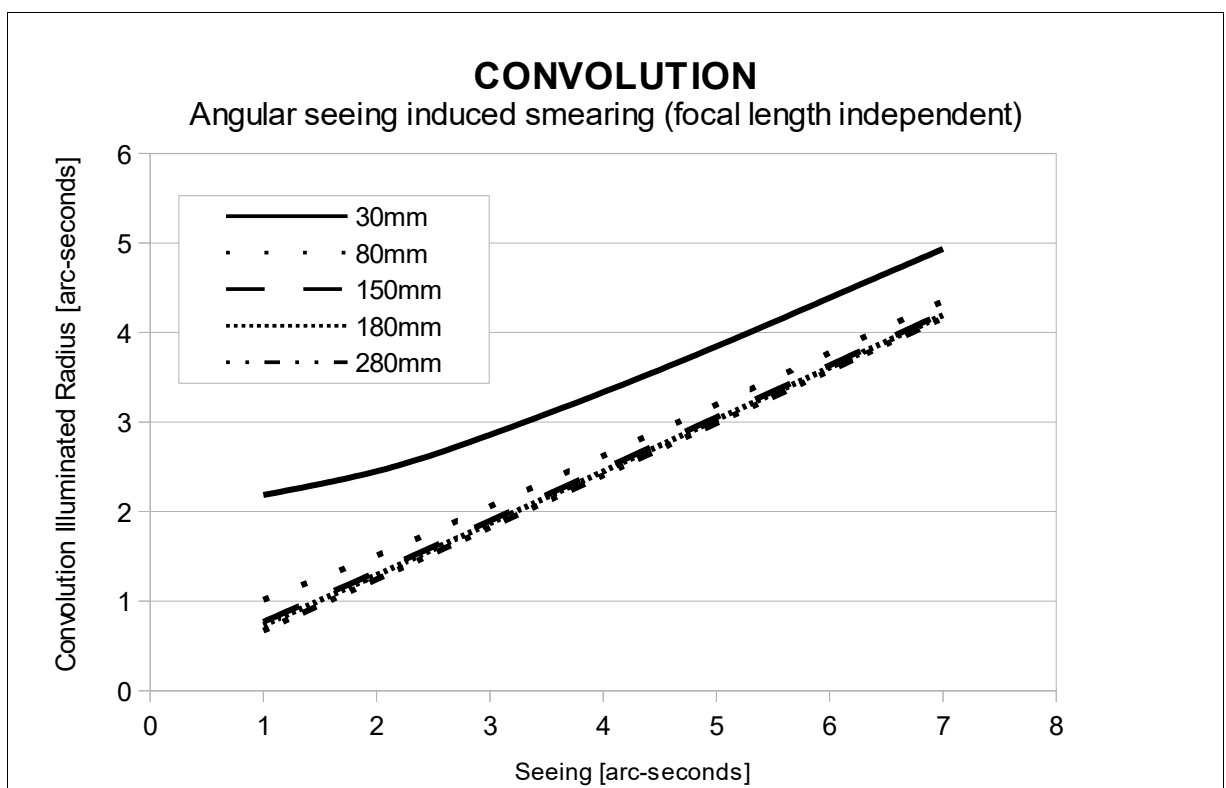

*Figure 4: The relationship between seeing and angular smearing by convolution.*

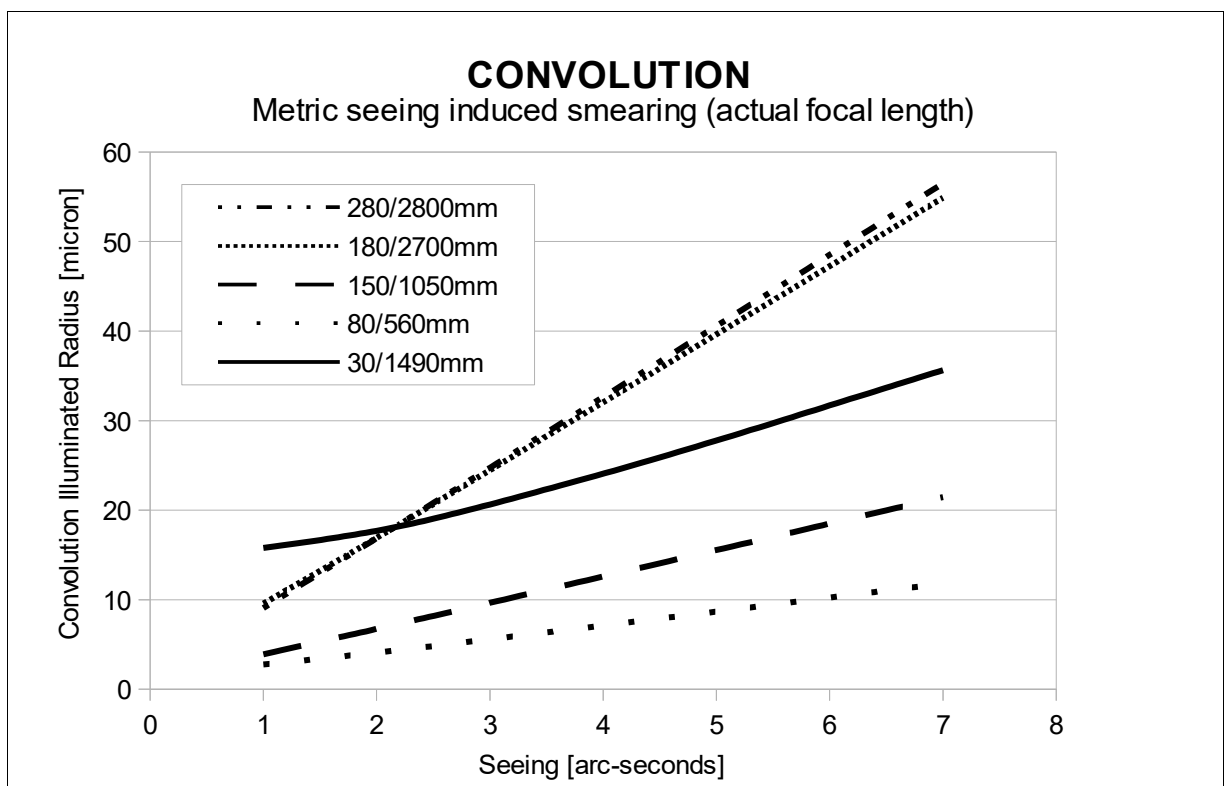

*Figure 5: The relationship between seeing and metric smearing by convolution.*

Figure 5 shows that the smaller aperture telescopes are capable of projecting a metrically smaller convoluted area at seeing above 2 arc-seconds. This seems to indicate that we can see finer details with them. However, the image size produced by these scopes depends on their focal lengths.

**Software implementation and validation**
In order to get an idea of what can actually be observed or imaged with these telescopes, software was written in PHP using above algorithms. The software takes a base-image as source and applies a scalable convolution to it. This base-image mimics an image taken with a camera behind a telescope (focal imaging) that is not affected by diffraction and seeing. In order to validate the software an image with only single light sources, each being one pixel in size, was used as input (see figure 6, left hand image). On the image two series of light sources can be discerned. They appear in pairs, at the left horizontally grouped, at the right at a 45 degrees angle. At the left their mutual distance starts at 1 pixel at the lower left, increasing with 1 pixel for each consecutive group going upwards and to the right. At the right the groups start with a mutual distance of √2 and increase by that amount for each additional group. At the centre of the image a single isolated light source is placed.

Processing for Fraunhofer diffraction alone results in a diffracted version of the image (see figure 6, central image). Due to the linear grey-scale only the Airy-disks are visible, the diffraction rings become visible after stretching the image in a non-linear way (see figure 6, right hand image). Processing this 516 × 418 pixel image took about 25 minutes on a modern computer, which may explain why Schaefer found it in practice difficult to calculate the convolution back in 1993.

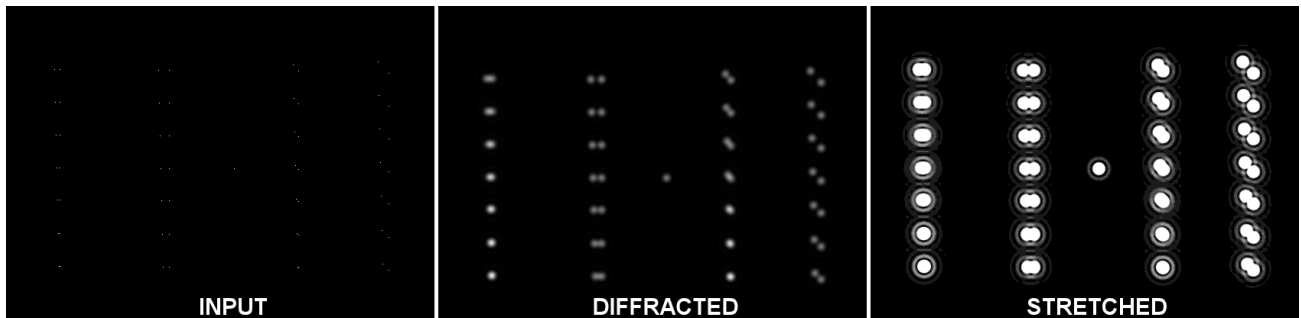

*Figure 6: Image used to test the smearing algorithm. At the left the input for the software, at the centre the resulting diffracted image, at the right the diffracted image stretched to show the diffraction rings.*

The processing was done using a setting of 10 pixels per Airy-disk radius. The resulting first minimum of the Fraunhofer diffraction should thus be 20 pixels in diameter, which indeed was confirmed using the stretched image. In addition the grouped light sources can be compared to the Rayleigh, Dawes and Sparrow criterion for detecting double stars (see figure 7). Rayleigh's criterion (1.00 × Airy-disk radius) should still show a significant intensity-dip halfway the two light-sources. Dawes' criterion (0.85 × Airy-disk radius) should have a minor dip, while Sparrow's criterion (0.77 × Airy-disk radius) should show no dip at all.

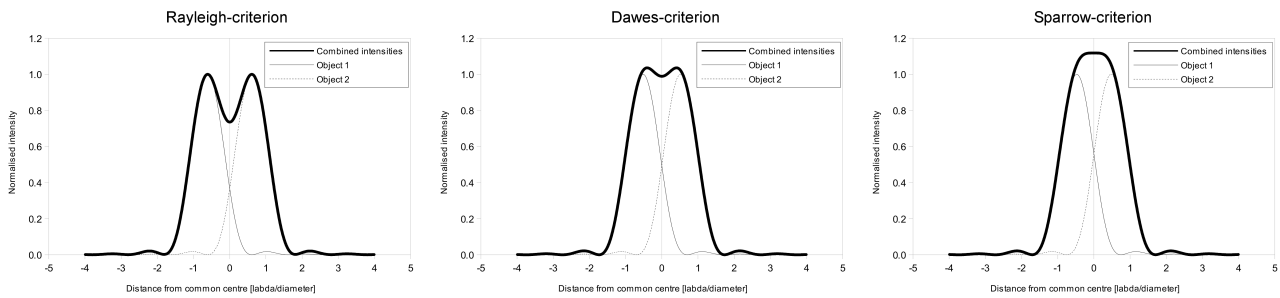

*Figure 7: Rayleigh, Dawes and Sparrow criterion.*

In the test image the Rayleigh criterion should thus be met at 10 pixels mutual distance, the Sparrow criterion at 8.5 pixels, while the Sparrow criterion should be at 7.7 pixels mutual distance. Figure 8 shows a four times enlarged version of the Fraunhofer diffracted test-image and careful measurements, done with image-processing software, confirmed that the diffraction was properly applied.[13]

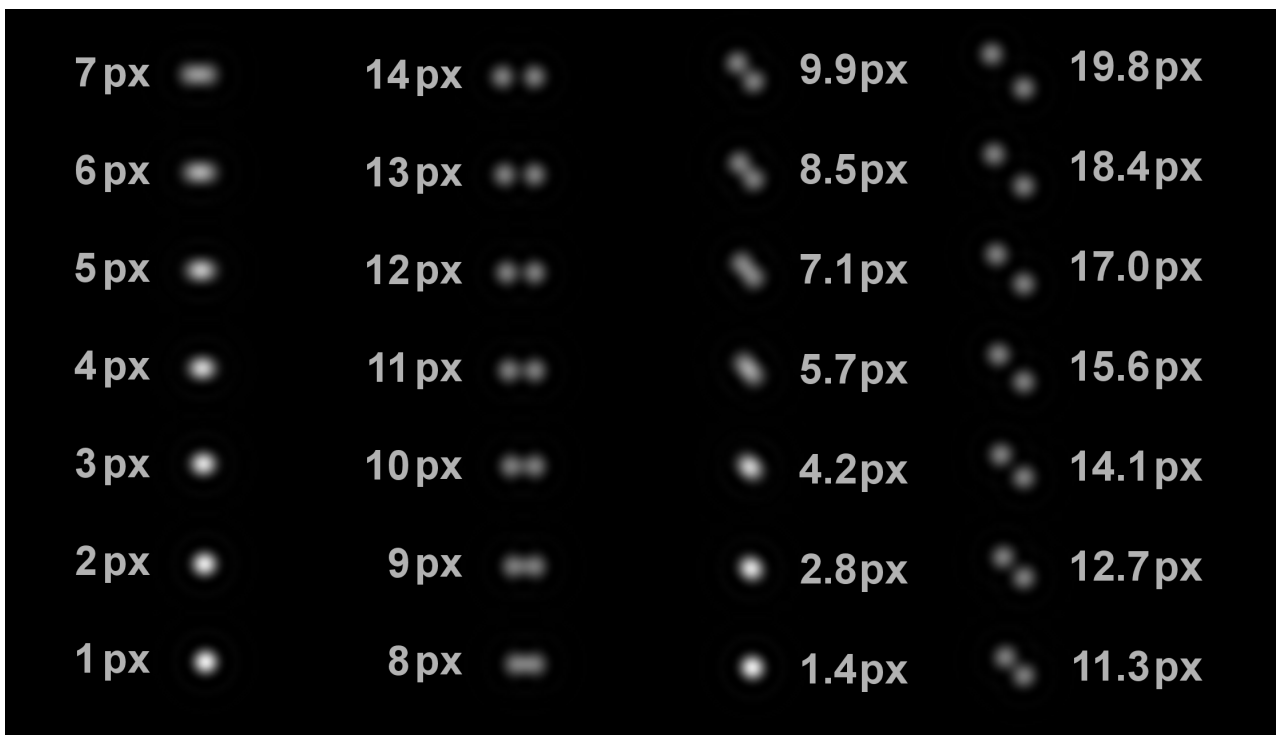

*Figure 8: Fraunhofer diffracted test image four times enlarged.*

A final validation was done using a single light source that was subjected to a convolution of diffraction and seeing with standard deviations $\sigma$ of respectively 0.0, 1.0 and 1.5 (see figure 9). These images were produced at a pixel scale of 100 pixels per Airy-disk radius and given a non-linear stretch to show the diffraction rings. As expected the diffraction rings are still visible when a $\sigma$ of 1.0 is applied (figure 9, central image), but gone at a $\sigma$ of 1.5 (figure 9, right hand image).[14]

---

13  The resizing has been done with Paint Shop Pro 22.2.0.8 x64 using bicubic resize, the measurements were done on the original image

14  With the diffraction rings gone it can be argued that at this $\sigma$ a telescope is seeing-limited. From [4] and figure 2 it can be gleaned that only small diameter telescopes under very good seeing conditions are truly diffraction limited.

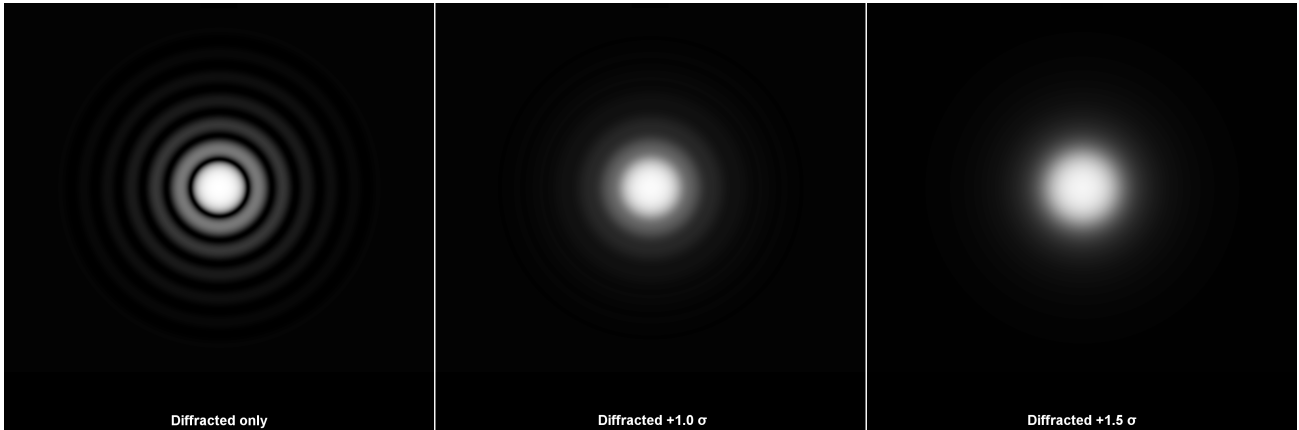

*Figure 9: A single light source subjected to a convolution of diffraction and seeing with standard deviations σ of respectively 0.0, 1.0 and 1.5.*

**Convoluting sunspot images**

With the software validated, a series of simulations could be created using a base-image with sunspots. This base-image (see figure 10) is a section of active region AR13057, taken on 17 July 2022 at around 08:49 UTC by one of the authors using a Schmidt-Cassegrain at f/20 (aperture 11 inch, 279mm, focal length 5588mm).[15] It shows a major sunspot with umbra and penumbra at the right and a smaller sunspot with umbra and penumbra at the left that is split by a light bridge. Left of that a pore and several micro-pores can be seen. Another micro pore is visible left of the largest sunspot and the background clearly shows granulation.

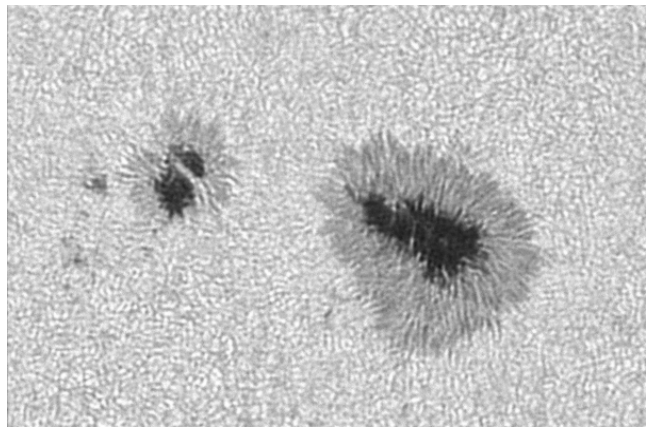

*Figure 10: Part of AR13057 as imaged on 17 July 2022. The penumbra of the left sunspot is approximately the size of the Earth.*

The image was processed using stacking and applying unsharp-mask, Lucy-Richardson deconvolution, contrast enhancement and additional sharpening.[16] It was then copied and scaled to half the resolution to mimic a photograph of that area taken with the same telescope at f/10, using a 2.95 micron pixel-size camera, but not affected by the smearing functions.[17] Copies of this image were made and resized to match the focal lengths of the other four telescopes and finally the theoretical resolution in number of pixels per Airy-disk radius was calculated for each of them to serve as input for the software.[17]

---

15 The telescope is a Celestron C11 EdgeHD, equipped with a TeleVue 2× PowerMate Barlow, Baader ND5.0 foil filter, Baader Continuum filter and ZWO ASI174MM monochrome camera with a pixel size of 5.9 micron.

16 Unsharp-mask, contrast enhancement and additional sharpening has been done with Paint Shop Pro 22.2.0.8 x64. The Lucy_Richardson deconvolution was done with ImPPG 0.5.4 (2019-02-02).

17 The resizing has been done with Paint Shop Pro 22.2.0.8 x64 using bicubic resize. We realise that this image still is affected by smearing, but it serves to show how it is affected by the convolution of diffraction and seeing.

Figure 11 shows the scaled images in the first row. These are the images that we expect to get if it were not for the smearing effects. Clearly the effect of focal length can be seen: the image and sunspot sizes are proportional with focal length. It are these images that were used in the software and to which the convolution at various seeing levels was applied. But in order to appreciate the differences between them, the images are shown at uniform size as in the second row, for which reason they were resized (figure 11, second row).[17]

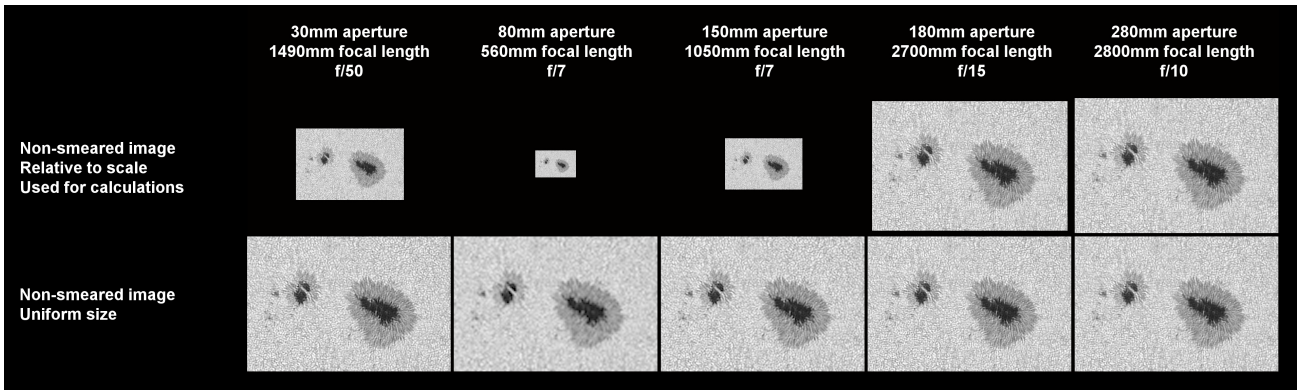

*Figure 11: The base-images for the convolution calculations.*

As the same theoretical camera was used for all five telescopes already differences can be seen between the images, the detail level being proportional to focal length.

The effects of the Fraunhofer diffraction alone (i.e. at a seeing of zero arc-seconds) is shown in figure 12. The first row shows the non-smeared images, while the second row shows the same images, but now with only Fraunhofer diffraction applied.[18]

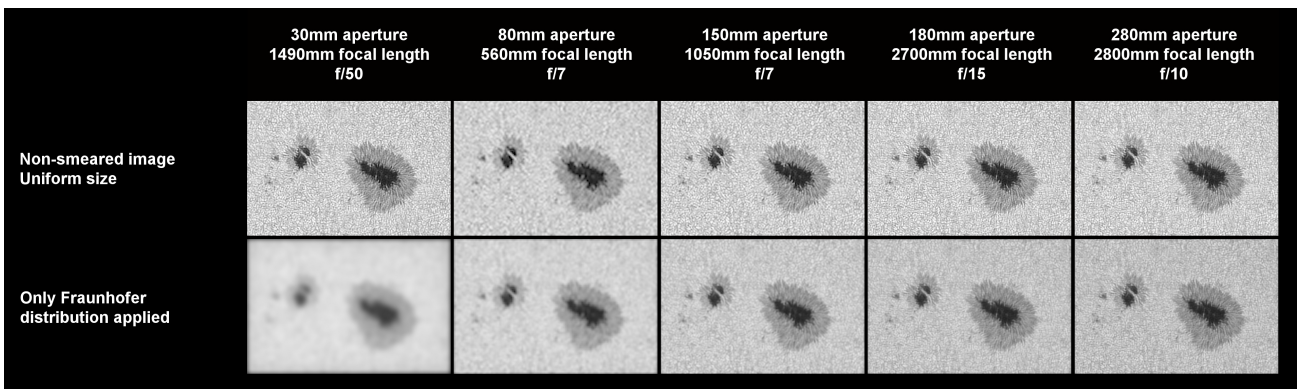

*Figure 12: The effect of Fraunhofer diffraction alone.*

It immediately becomes clear that the small aperture of the Galilean type telescope has a significant effect on the detail level: the granulation is gone and the micro pores have merged into a single pore. The same, but to a lesser extent, applies to the 80mm aperture f/7 telescope. The larger aperture telescopes are only marginally affected, but still the effect of smearing is visible.

---

18  In all simulations the histogram has been normalised to 90% of the maximum image intensity level of 255.

In figure 13 the combined effect of Fraunhofer diffraction and seeing is shown, calculated using the convolution discussed in this article. Most obvious are the first and last column, the extremes when it comes to aperture. The first column shows the Galilean type telescope of 30mm aperture and 1490mm focal length, the last a 280mm aperture, 2800mm focal length telescope.

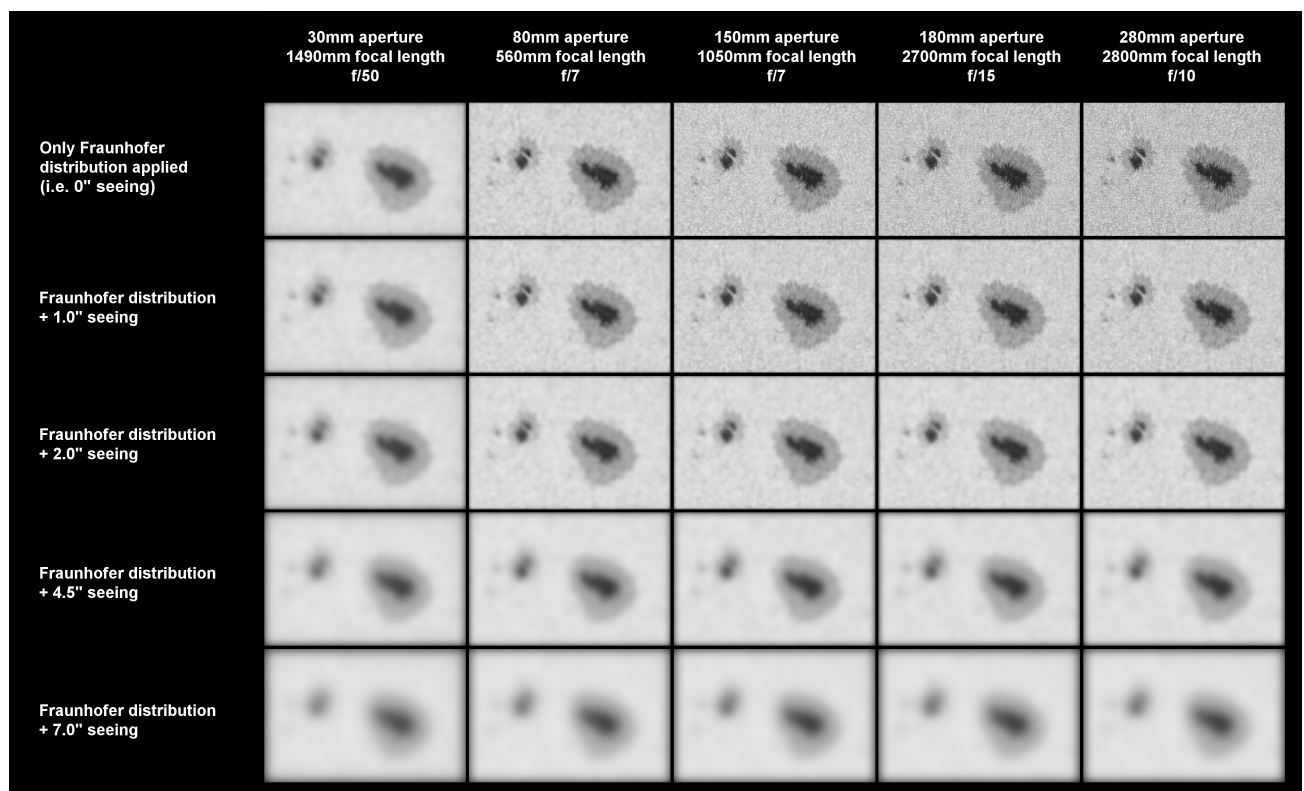

*Figure 13: The effect of convolution under different seeing conditions.*

The 30mm aperture telescope only suffers marginally from deterioration with worsening seeing. The image-detail is relatively poor from the onset as this telescope is diffraction limited. Still the image is clearly affected by it. Significant is the loss of detail in the sunspot with the light bridge. At 1 and 2 arc-seconds seeing, the light bridge is still visible, resulting in a count of 2 sunspots, but at 4.5 arc-seconds seeing the light bridge becomes vague, while at 7 arc-seconds it totally disappears. Likewise the micro pores at the left join to form a single pore straight from the best seeing, but they almost disappear when the seeing worsens to 7 arc-seconds. Therefore, Schaefer's remark that "...small telescopes (or aperture stops) with *D* less than 5 cm or so will remove the effect of variable seeing" cannot be justified.

Larger aperture telescopes are heavily affected by seeing as well, but comparing the five telescopes at the same seeing-level, it becomes clear that a larger aperture telescope always produces an equally or more detailed image than a smaller aperture telescope. The 280mm aperture telescope still shows the light bridge at 4.5 arc-seconds seeing and the micro pores, although merged into a single pore, are still visible at a seeing of 7 arc-seconds. None of the simulations indicate that a larger aperture telescope performs worse than a smaller one under the same seeing conditions. Even at a seeing level of 7 arc-seconds the largest diameter telescope in the test still performs better than the smallest one, albeit marginally.

**Comparison with practice**

The above simulation shows that there still is a noticeable effect of the seeing when using a 30mm telescope. Daily observations made since March 2021 by one of the authors (NdH) confirm that this indeed is the case. Small objects like the pores and micro pores tend to (dis)appear continuously with seeing variation, similar to what is shown in above simulations.

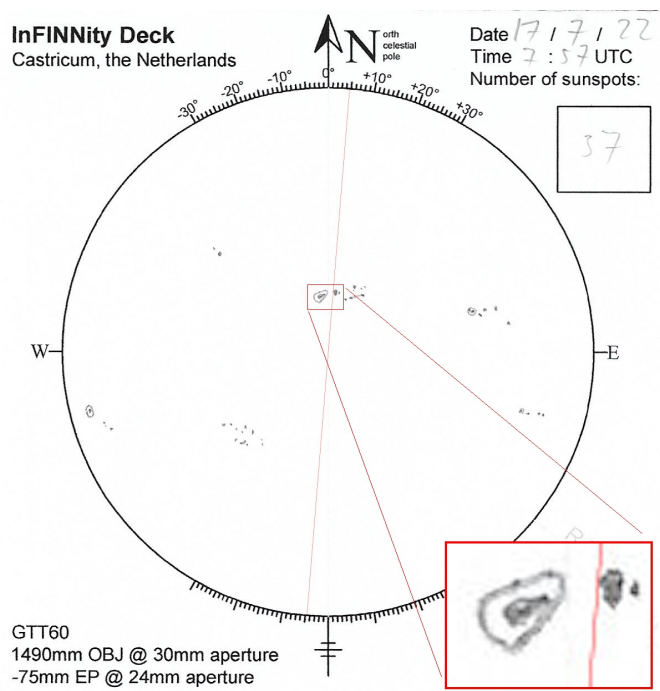

But even under the conditions as on 17 July 2022 (average seeing was 3.5 - 4.0 arc-seconds), these minor sunspots are hard to see, which is partially caused by the size at which they are projected and partially due to the lack of contrast due to the smearing. Figure 14 shows the sketch by the author for that day. The Sun is projected within an 80mm diameter circle. The simulated images shown in figures 11 to 13 cover the area marked with the red rectangle and measures only 6 by 4 four millimetres. Even though that at the seeing level of that day

*Figure 14: Sketch of the Sun made using a 30mm Galilean type telescope. The approximately 6 × 4 mm area of the simulation is marked with a red rectangle.*

both the pore and the group of micro pores may have been visible, only the pore was actually drawn.

To get an idea how seeing really affects our observations we have to look at how figure 10 was created under these circumstances. A large aperture telescope, like the 11" SCT used for the image, shows distinct variation in image detail under varying seeing and that is exactly what is noticed when observing or imaging with it. Images as in figure 10 can only be achieved by a process called *lucky imaging*. In that process a large number of images (in this case 500) are taken with an as short as possible exposure time. The whole process is triggered by a *Solar Scintillation Seeing Monitor* (SSSM), which starts the capture when seeing is under a pre-set threshold.[19] That the seeing was reported as being 3.5 to 4 arc-seconds on average means that there were periods with better and worse seeing.

Figure 15 is an example of the SSSM data, recorded on a different date with typical seeing conditions for the Netherlands. The average seeing (red line in the graph) varies between 3.5 and 6.0 arc-seconds, but the best seeing that day was between 2.0 and 2.5 arc-seconds and those would be the moments that the finest details would become visible. The dips in the yellow curve indicate passing clouds.

---

19 The SSSM reports seeing in arc-seconds of diameter. The source code of the Arduino software reads `variationValue = (4.46 * sqrt(variationValue / numSamples) + variationOffset) / intensityValue; // Find RMS`. The factor 4.46 is the result of the circuitry magnifying ΔIntensity by a factor of 20M/47k ≈ 425.5. With the Sun being approximately 1900" in diameter the factor 4.46 follows from 1900/425.5 and thus the SSSM reports seeing as the diameter of it, see *Solar Chat Forum*, "DIY Solar Scintillation Seeing Monitor with FC support" (Sat Jun 17, 2017 3:35 am), url: https://solarchatforum.com/viewtopic.php?f=9&t=16746&hilit=ssm+diy&start=25#p206567 (last accessed 28 July 2022).

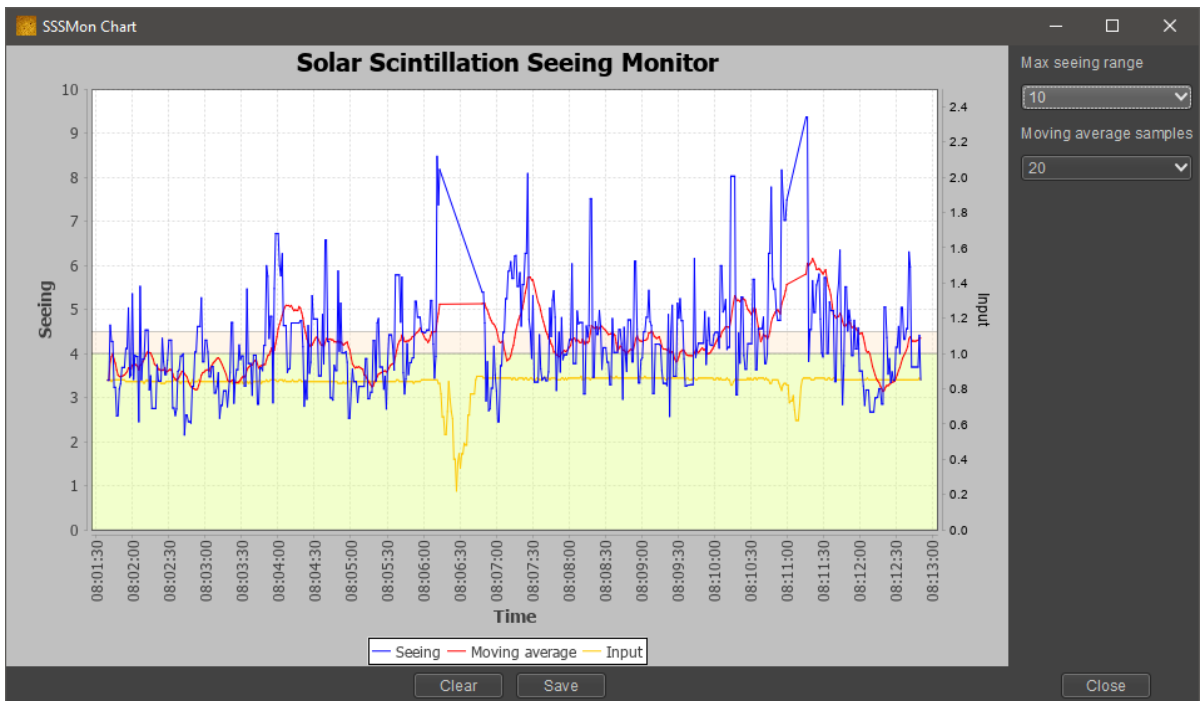

*Figure 15: Data recorded by the Solar Scintillation Seeing Monitor.*

The SSSM waits until those better seeing-periods to start the capture (in this case the trigger level was set to 4.0 arc seconds). Using dedicated software only the best of the recorded images are stacked to improve signal to noise ratio, after which the image is being processed to the result shown in figure 10.

In this respect sketching sunspots is quite similar to lucky imaging. The projected image is analysed for a prolonged period of time and each time a sunspot is noticed it is drawn onto the sheet. So even when a sunspot is appearing and disappearing continuously or only appearing every now and then, it will be recorded. When a SSSM or other seeing measuring device is available to indicate seeing levels it is better to note down the lowest seeing level rather than the average as it is that best seeing level that determines what can and cannot be seen.

Although even small aperture telescopes are affected by seeing, due to the 'lucky imaging'-way of observing the resulting sketches are hardly affected. Figure 16 shows the percentage of sunspot-number found by the author using a 30mm aperture Galilean type telescope versus the average recorded seeing in 87 observations in the period from 3 September 2021 to 29 April 2022. As reference the average sunspot count $R$ was used from observatories in Brussels, Kanzelhohe and Locarno ($R$ = groups × 10 + sunspots). The sunspots were counted using the same grouping method as at the reference observatories.

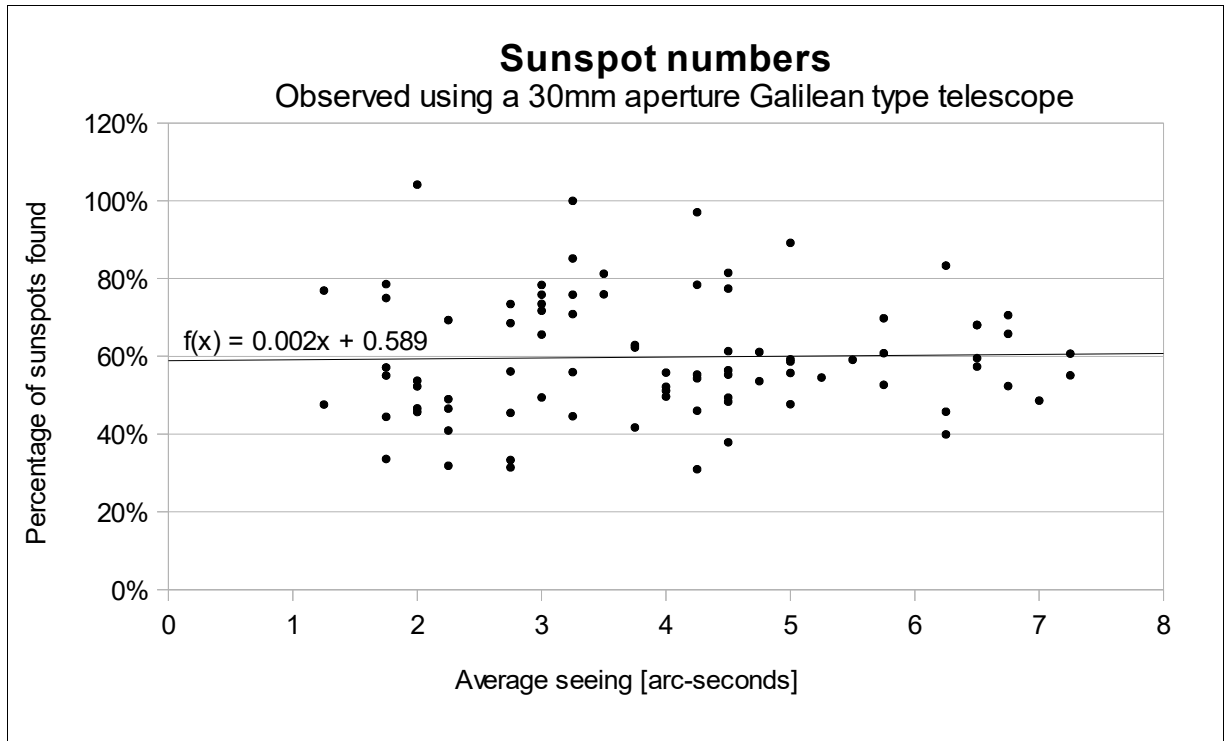

*Figure 16: Percentage of sunspot numbers found using a 30mm Galilean type telescope.*

Only a very small correlation of about 0.2% per arc-second of seeing can be found from this data, indicating that, although the small aperture telescope is affected by seeing, the seeing does not play a significant role in the number of sunspots and groups found at this aperture. Sadly enough only the average seeing has been recorded, future observations with the minimum seeing recorded may show if this correlation still remains this small.

**Conclusion**

Using convolution the effect of atmospheric seeing and telescope aperture has been analysed. Although seeing affects larger aperture telescopes more than smaller aperture ones, the effect of seeing remains visible in telescopes with apertures as small as 30mm. At larger apertures the effect of seeing is more pronounced, but at the same time larger aperture telescopes still preform (sometimes only marginally) better than smaller aperture telescopes under the same conditions. Seeing seems to have only little effect on sunspot counts when using a small aperture telescope, but that is mainly due to the way we observe and not so much due to the effect of seeing itself. More data is required, however, to further substantiate this. When recording seeing values together with sunspot numbers it is recommended to note down the lowest observed seeing level instead of the average, as it is the lowest level that determines what details can be observed.